\documentclass[a4paper]{article}
\pdfoutput=1
\usepackage[a4paper]{geometry}
\usepackage{amsmath}
\usepackage{color}
\usepackage{graphicx}
\usepackage{caption}
\usepackage{subcaption}
\usepackage{tabularx}
\usepackage{bigstrut}
\usepackage{multirow}
\usepackage{authblk}
\usepackage[utf8]{inputenc}
\usepackage[pdfauthor={Declan Millar}, pdftitle={Using asymmetry observables to discover and distinguish Z' signals in top pair production with the lepton-plus-jets final state at the LHC}, pdfkeywords={Z' zprime boson top quark pair production charge spin asymmetry asymmetries lepton-plus-jets},hidelinks]{hyperref}
\usepackage[numbers,compress]{natbib}
\usepackage{float}

\title{Using asymmetry observables to discover and distinguish \texorpdfstring{$Z'$}{Z'} signals in top pair production with the lepton-plus-jets final state at the LHC}

\author[1]{Lucio Cerrito\thanks{E-mail: {\tt lucio.cerrito@cern.ch}}}
\author[2,3]{Declan Millar\thanks{E-mail: {\tt declan.millar@cern.ch }}}
\author[3]{Stefano Moretti\thanks{E-mail: {\tt S.Moretti@soton.ac.uk}}} 
\author[4]{Francesco Span\`{o}\thanks{E-mail: {\tt francesco.spano@cern.ch}}}

\affil[1]{\it Department of Physics, University of Rome Tor Vergata and INFN, Via Della Ricerca Scientifica, 1, Rome, 00133, Italy}
\affil[2]{\it School of Physics and Astronomy, Queen Mary University of London, Mile End Road, London E1 4NS, United Kingdom}
\affil[3]{\it School of Physics and Astronomy, University of Southampton, Highfield, Southampton SO17 1BJ, United Kingdom}
\affil[4]{\it Department of Physics, Royal Holloway University of London, Egham Hill, Egham TW20 0EX, United Kingdom}

\begin{document}

\pagenumbering{Alph}
 
\begin{titlepage}

  \maketitle

  \begin{abstract}
    We study the sensitivity of top pair production with six-fermion decay at the LHC to the presence and nature of an underlying $Z'$ boson, accounting for full tree-level Standard Model $t\bar{t}$ interference, with all intermediate particles allowed off-shell. We concentrate on the lepton-plus-jets final state and simulate experimental conditions, including kinematic requirements and top quark pair reconstruction in the presence of missing transverse energy and combinatorial ambiguity in jet-top assignment. We focus on the differential mass spectra of the cross section and asymmetry observables, especially demonstrating the use of the latter in probing the coupling structure of a new neutral resonance, in addition to cases in which the asymmetry forms a complementary discovery observable.
  \end{abstract}

  \thispagestyle{empty}

\end{titlepage}

\pagenumbering{arabic}

\section{Introduction}
\label{sec:introduction}

New fundamental, massive, neutral, spin-1 gauge bosons ($Z'$) appear ubiquitously in theories Beyond the Standard Model (BSM). The strongest limits for such a state generally exist for the $e^+e^-$ and $\mu^+\mu^-$ signatures, known collectively as Drell-Yan (DY). However, in addition to their importance in extracting the couplings to top quarks, resonance searches in the $t\bar{t}$ channel can offer additional handles on the properties of a $Z'$ due to uniquely available asymmetry observables, owing to the fact that (anti)tops decay prior to hadronisation and spin information is effectively transmitted to decay products. Their definition in $t\bar{t}$, however, requires the reconstruction of the top quark pair. In these proceedings we summarise our study of the sensitivity to the presence of a single $Z'$ boson at the Large Hadron Collider (LHC) arising from a number of generationally universal benchmark models (section~\ref{sec:models}), as presented in our recently submitted paper~\cite{cerrito2016}. We simulate top pair production and six-fermion decay mediated by a $Z'$ with full tree-level SM interference and all intermediate particles allowed off-shell, with analysis focused on the lepton-plus-jets final state, and imitating some experimental conditions at the parton level (section~\ref{sec:method}). We assess the prospect for an LHC analysis to profile a $Z'$ boson mediating $t\bar{t}$ production, using the cross section in combination with asymmetry observables, with results and conclusions in section~\ref{sec:results} and~\ref{sec:conclusions}, respectively.

\section{Models}
\label{sec:models}

There are several candidates for a Grand Unified Theory (GUT), a hypothetical enlarged gauge symmetry, motivated by gauge coupling unification at approximately the $10^{16}$ GeV energy scale. $Z'$ often arise due to the residual U$(1)$ gauge symmetries after their spontaneous symmetry breaking to the familiar SM gauge structure. We study a number of benchmark examples of such models. These may be classified into three types: $E_6$ inspired models, generalised Left-Right (GLR) symmetric models and General Sequential Models (GSMs)~\cite{accomando2011}.

One may propose that the gauge symmetry group at the GUT scale is E$_6$. When recovering the SM, two residual symmetries U$(1)_\psi$ and U$(1)_\chi$ emerge, which may survive down to the TeV scale. LR symmetric models introduce a new isospin group, SU$(2)_R$, perfectly analogous to the SU$(2)_L$ group of the SM, but which acts on right-handed fields. This symmetry may arise naturally when breaking an SO$(10)$ gauge symmetry. We are particularly interested in the residual U$(1)_R$ and U$(1)_{B-L}$ symmetries, where the former is related to $T^3_R$, and $B$ and $L$ refer to Baryon and Lepton number, respectively. An SSM $Z'$ has fermionic couplings identical to those of the SM $Z$ boson, but is generically heavier. In the SM the $Z$ couplings to fermions are uniquely determined by well defined eigenvalues of the $T^3_L$ and $Q$ generators, the third isospin component and the Electro-Magnetic (EM) charge.

For each class we may take a general linear combination of the appropriate operators and fix $g'$, varying the angular parameter dictating the relative strengths of the component generators, until we recover interesting limits. These models are all universal, with the same coupling strength to each generation of fermion. Therefore, as with an SSM $Z'$, the strongest experimental limits come from the DY channel. The limits for these models have been extracted based on DY results, at $\sqrt{s}=7$ and $8$~TeV with an integrated luminosity of $L=20$~fb$^{-1}$, from the CMS collaboration~\cite{thecmscollaboration2015}\nocite{theatlascollaboration2014c} by Accomando et al.~\cite{accomando2016}, with general consensus that such a state is excluded below $3$ TeV.


\section{Method}
\label{sec:method}

Measuring $\theta$ as the angle between the top and the incoming quark direction, in the parton centre of mass frame, we define the forward-backward asymmetry:
\begin{equation}
	A_{FB}=\frac{N_{t}(\cos\theta>0)-N_t(\cos\theta<0)}{N_t(\cos\theta>0)+N_t(\cos\theta<0)}, \quad \cos\theta^* =\frac{y_{tt}}{|y_{tt}|}\cos\theta 
\end{equation}
With hadrons in the initial state, the quark direction is indeterminate. However, the $q$ is likely to carry a larger partonic momentum fraction $x$ than the $\bar{q}$ in $\bar{x}$. Therefore, to define $A^{*}_{FB}$ we choose the $z^*$ axis to lie along the boost direction. The top polarisation asymmetry ($A_{L}$), measures the net polarisation of the (anti)top quark by subtracting events with positive and negative helicities:
\begin{equation}
  A_{L}=\frac{N(+,+)+N(+,-)-N(-,-)-N(-,+)}{N(+,+)+N(+,-)+N(-,-)+N(-,+)}, \quad \frac{1}{\Gamma_l}\frac{d\Gamma_l}{dcos\theta_l}=\frac{1}{2}(1 + A_L \cos\theta_l),
\end{equation}
where $\lambda_{t}$($\lambda_{\bar{t}}$) denote the eigenvalues under the helicity operator of $t$($\bar{t}$). Information about the top spin is preserved in the distribution of $\cos\theta_l$. We construct two dimensional histograms in $m_{tt}$ and $(\cos\theta_{l})$, and equate the gradient of a fitted straight line to $A_{L}$.

In each of the models, the residual U$(1)'$ gauge symmetry is broken around the TeV scale, resulting in a massive $Z'$ boson. This leads to an additional term in the low-energy Lagrangian, from which we may calculate the unique $Z'$ coupling structure for each observable:
\begin{align}
  \mathcal{L} &\supset g^\prime Z^\prime_\mu \bar{\psi}_f\gamma^\mu(f_V - f_A\gamma_5)\psi_f,
  \label{eq:zprime_lagrangian}\\
  \hat{\sigma} &\propto \left(q_V^2 + q_A^2\right)\left((4 - \beta^2)t_V^2 + t_A^2\right),\\
  A_{FB} &\propto q_V q_A t_V t_A,\\
  A_{L} &\propto \left(q_V^2 + q_A^2\right)t_V t_A,
\end{align}
where $f_V$ and $f_A$ are the vector and axial-vector couplings of a specific fermion ($f$).\nocite{hagiwara1992,stelzer1994,lepage1978}

While a parton-level analysis, we incorporate restraints encountered with reconstructed data, to assess, in a preliminary way, whether these observables survive. The collider signature for our process is a single $e$ or $\mu$ produced with at least four jets, in addition to missing transverse energy ($E^{\rm miss}_{T}$). Experimentally, the $b$-tagged jet charge is indeterminate and there is ambiguity in $b$-jet (anti)top assignment. We solely identify $E^{\rm miss}_{T}$ with the transverse neutrino momentum. Assuming an on-shell $W^\pm$ we may find approximate solutions for the longitudinal component of the neutrino momentum as the roots of a quadratic equation. In order to reconstruct the event, we account for bottom-top assignment and $p_z^\nu$ solution selection simultaneously, using a chi-square-like test, by minimising the variable $\chi^2$:
\begin{equation}
  \chi^2 = \left(\frac{m_{bl\nu}-m_{t}}{\Gamma_t}\right)^2 + \left(\frac{m_{bqq}-m_{t}}{\Gamma_t}\right)^2,
  \label{eq:chi2}
\end{equation}
where $m_{bl\nu}$ and $m_{bqq}$ are the invariant mass of the leptonic and hadronic (anti)top, respectively.

In order to characterise the sensitivity to each of these $Z'$ models, we test the null hypothesis, which includes only the known $t\bar{t}$ processes of the SM, assuming the alternative hypothesis ($H$), which includes the SM processes with the addition of a single $Z'$, using the profile Likelihood ratio as a test statistic, approximated using the large sample limit, as described in~\cite{cowan2011}. This method is fully general for any $n$D histogram, and we test both $1$D histograms in $m_{tt}$, and $2$D in $m_{tt}$ and the defining variable of each asymmetry to assess their combined significance.

\section{Results}
\label{sec:results}

\begin{figure}[H]
  \centering
  \begin{subfigure}{0.494\textwidth}
    \includegraphics[width=\textwidth]{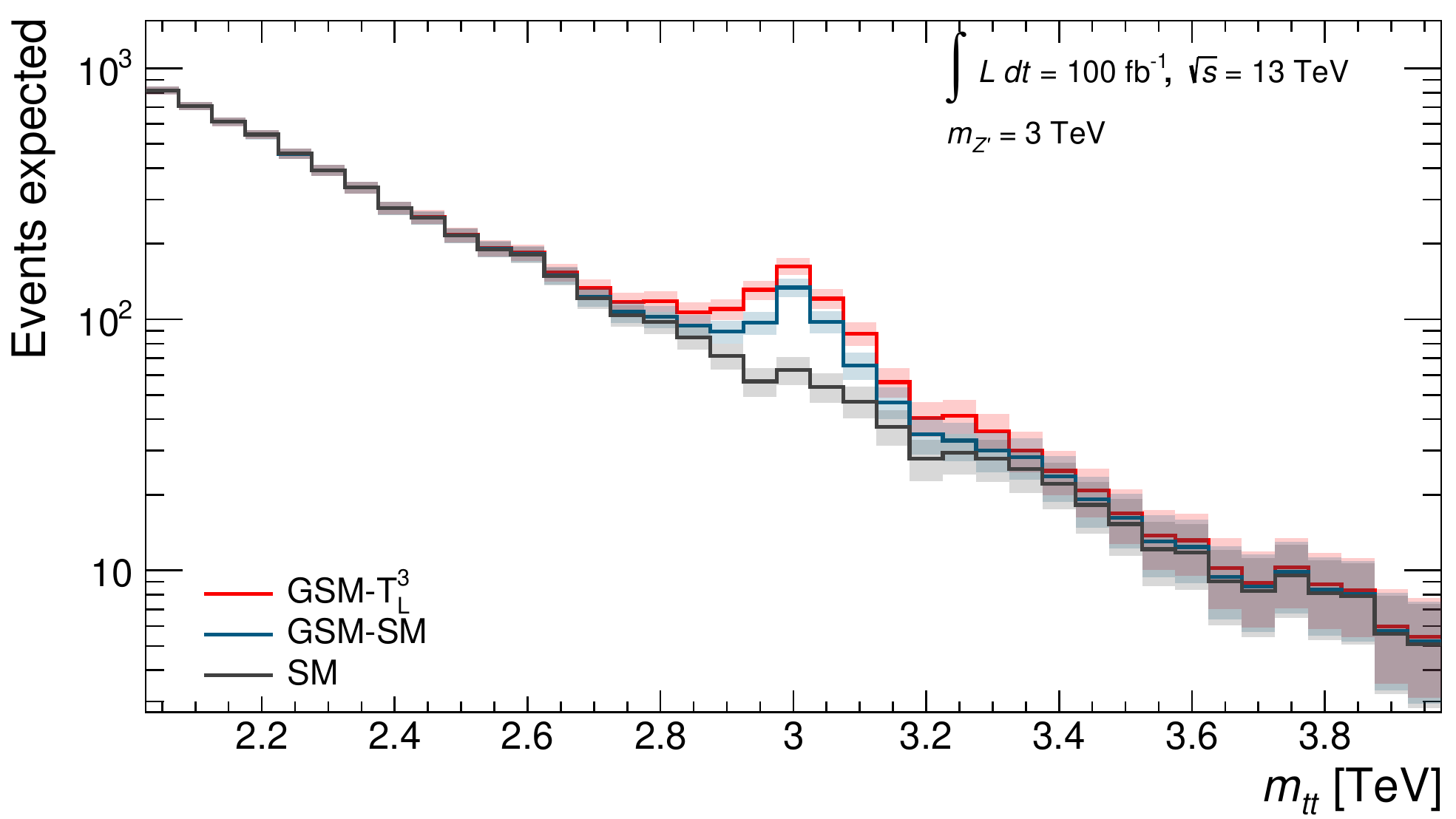}
    \caption{Events expected - GSM models}
  \end{subfigure}
  \begin{subfigure}{0.494\textwidth}
    \includegraphics[width=\textwidth]{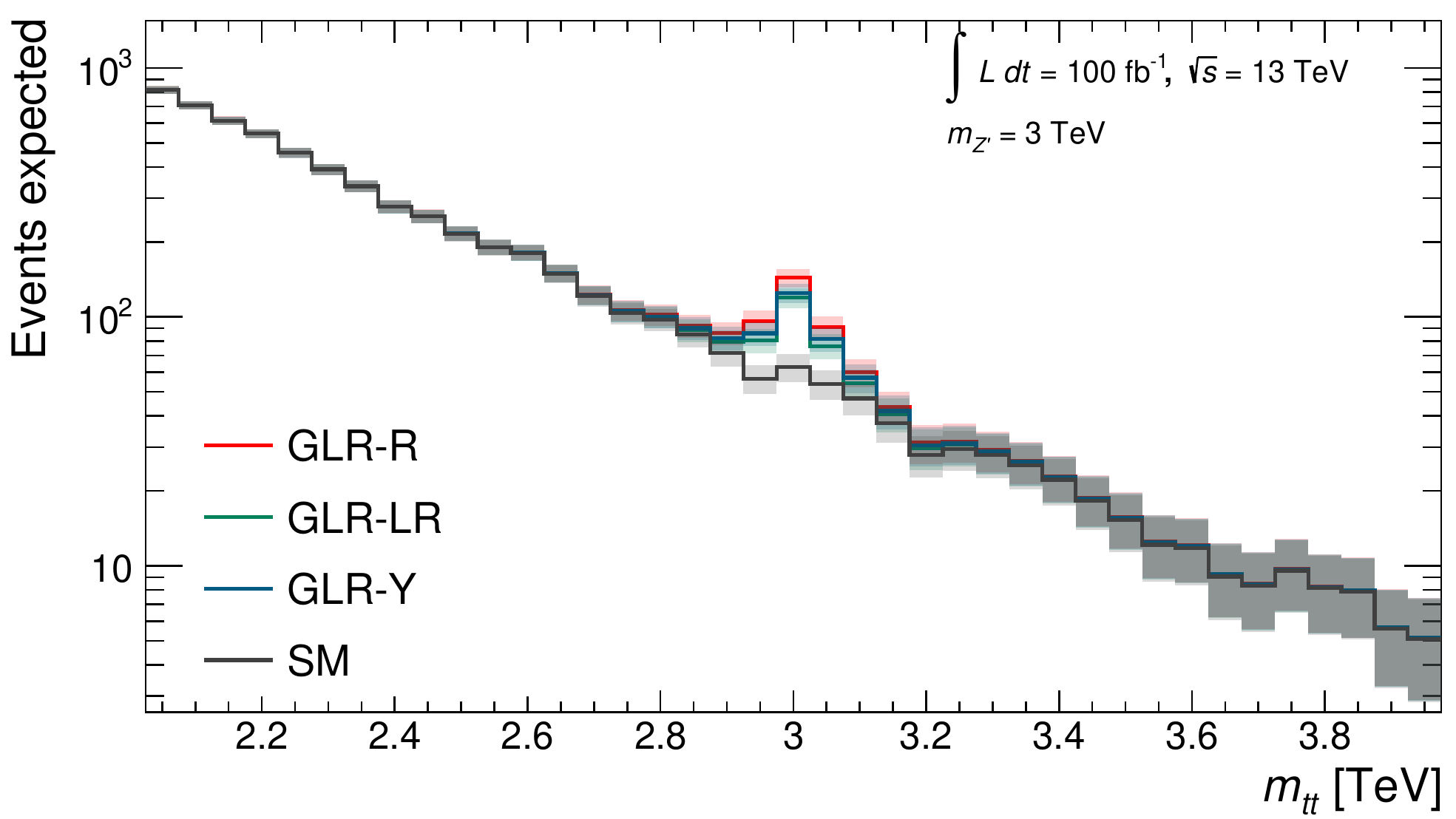}
    \caption{Events expected - GLR models}
  \end{subfigure}
  \begin{subfigure}{0.494\textwidth}
    \includegraphics[width=\textwidth]{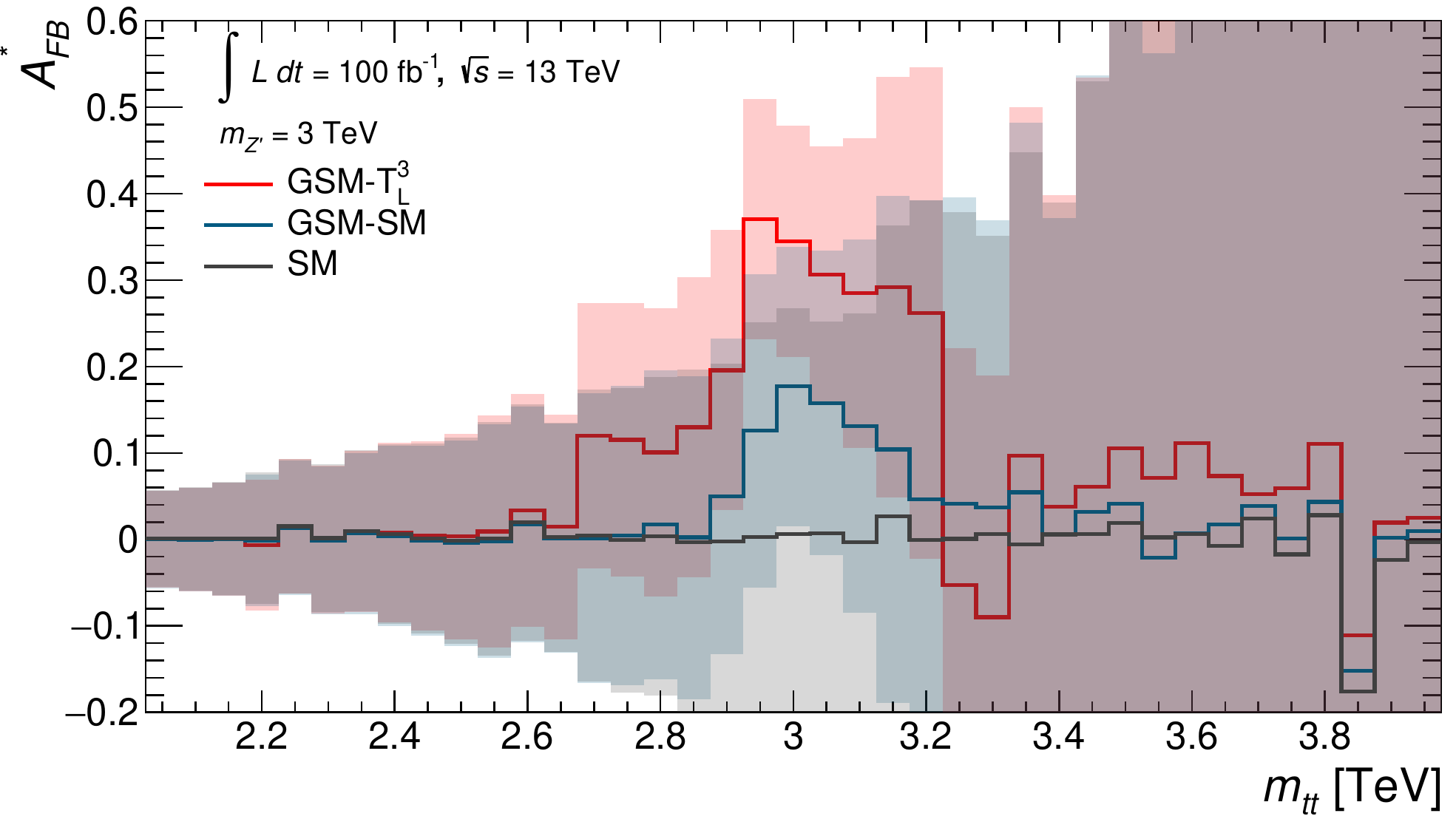}
    \caption{$A_{FB}^{*}$ - GSM models}
  \end{subfigure}
  \begin{subfigure}{0.494\textwidth}
    \includegraphics[width=\textwidth]{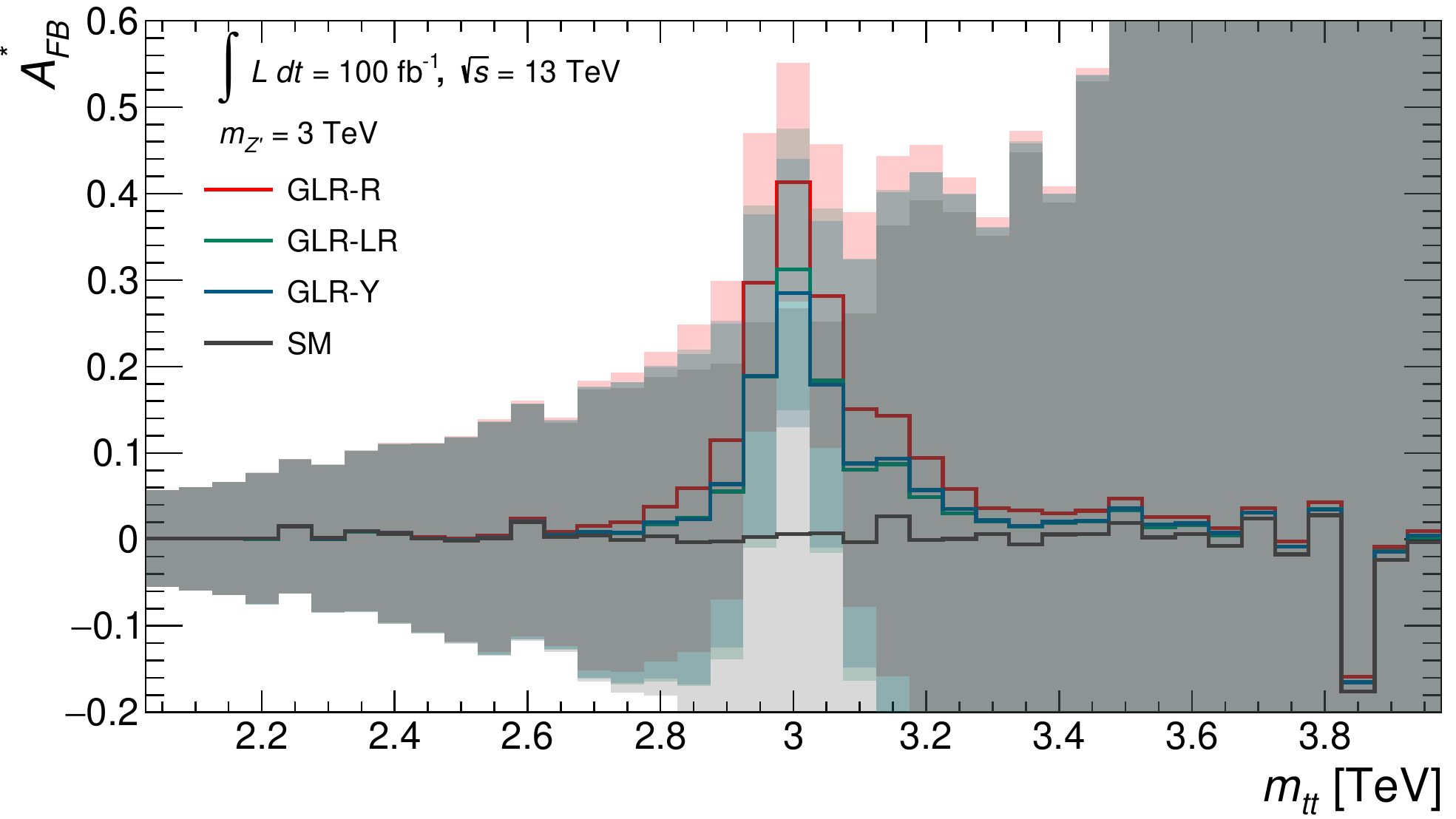}
    \caption{$A_{FB}^{*}$ - GLR models}
  \end{subfigure}
  \begin{subfigure}{0.494\textwidth}
    \includegraphics[width=\textwidth]{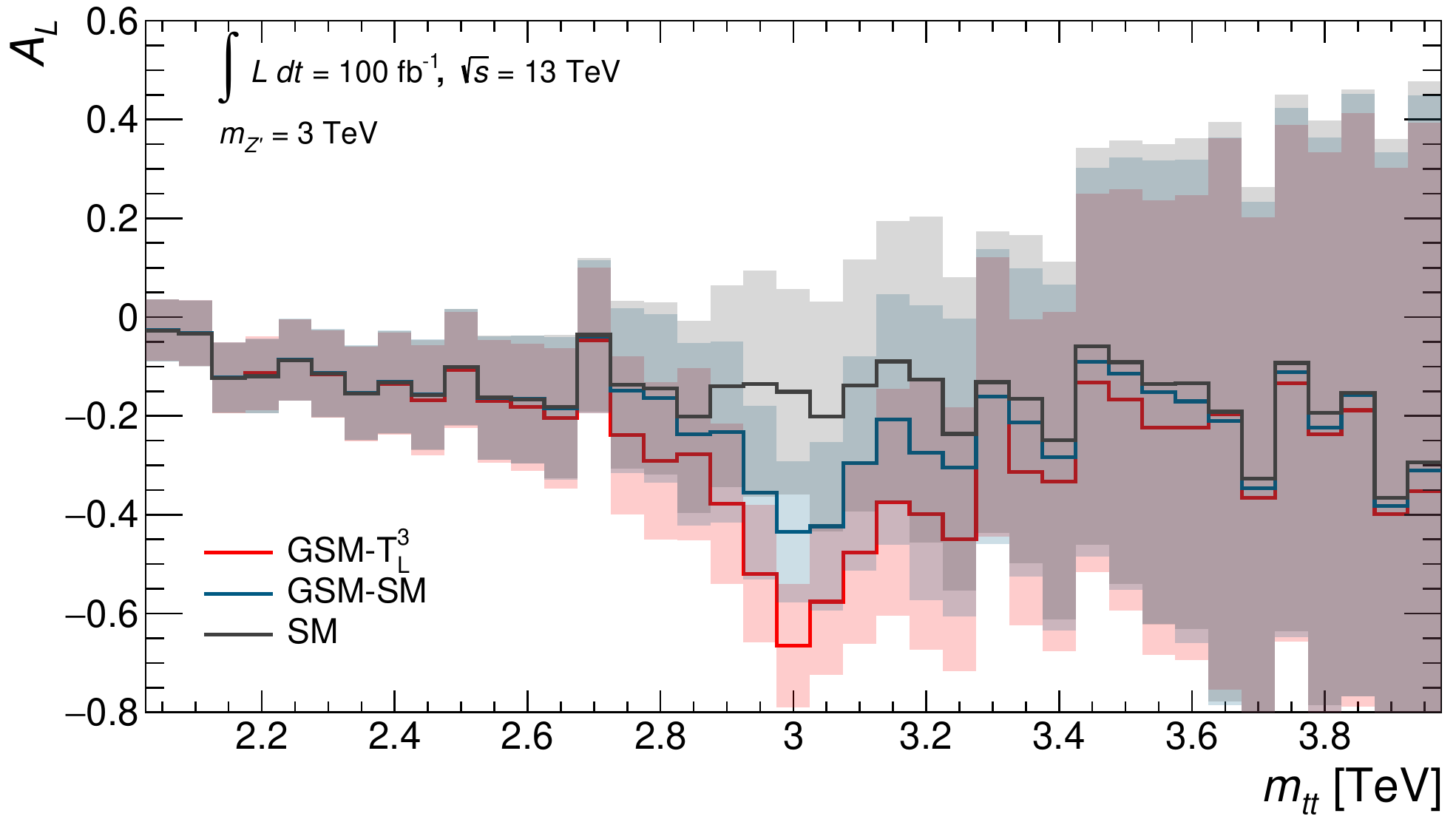}
    \caption{$A_L$ - GSM models}
  \end{subfigure}
  \begin{subfigure}{0.494\textwidth}
    \includegraphics[width=\textwidth]{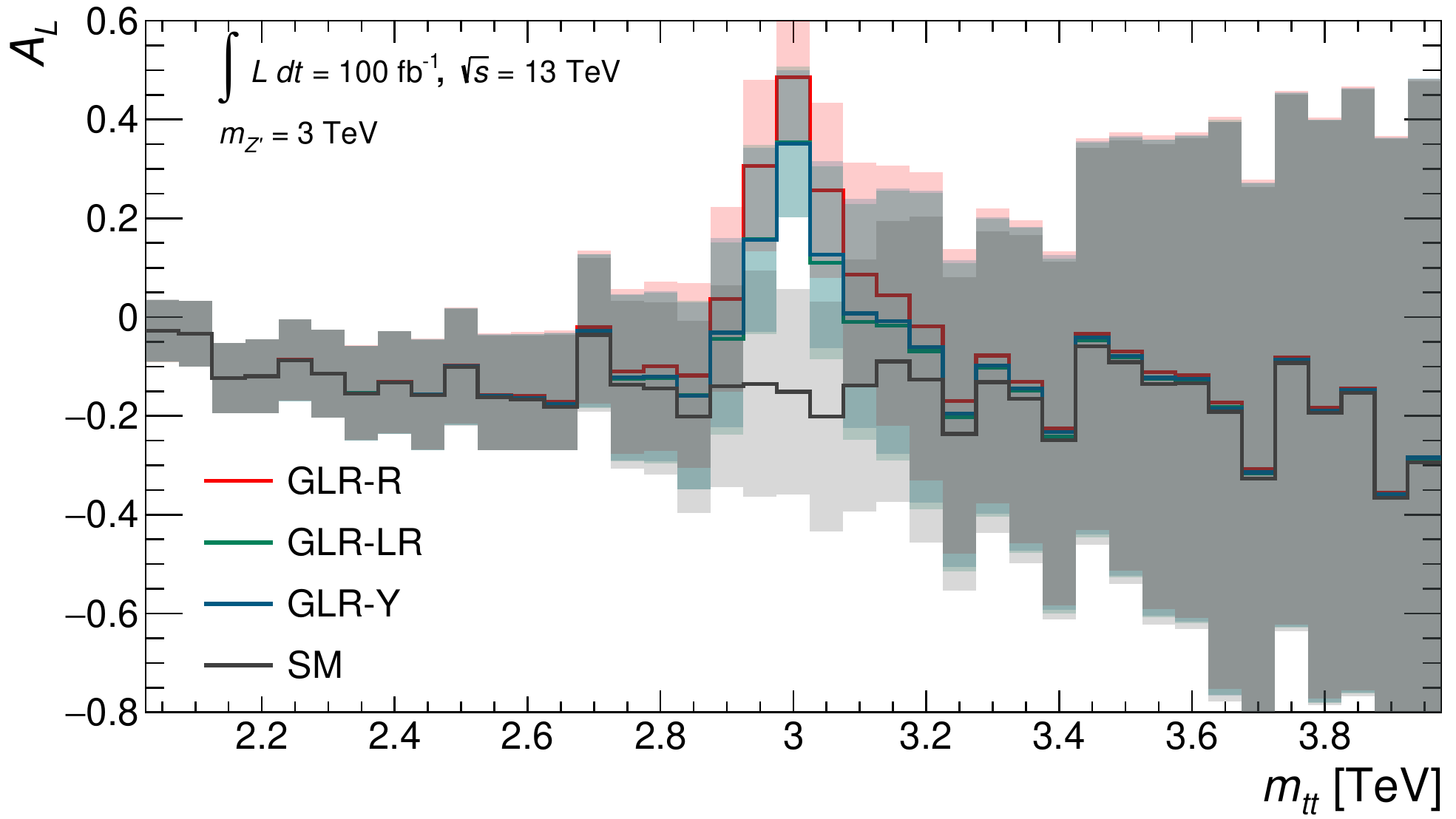}
    \caption{$A_L$ - GLR models}
  \end{subfigure}
  \caption{Expected distributions for each of our observables of interest, with an integrated luminosity of $100$~fb$^{-1}$, at $\sqrt{s}=13$~TeV. The shaded bands indicate the projected statistical uncertainty.}
  \label{fig:distinguishing}
\end{figure}

Figure~\ref{fig:distinguishing} shows plots for the differential cross section, $A_{FB}^{*}$ and $A_{L}$. The statistical error is quantified for this luminosity assuming Poisson errors. The absent models, including all of the $E_6$ class, only produce an asymmetry via the interference term, which generally gives an undetectable enhancement with respect to the SM yield. The absence of a corresponding peak in either asymmetry offers an additional handle on diagnosing a discovered $Z'$. The cross section, profiled in $m_{tt}$, shows a very visible peak for all models. The GSM models feature a greater peak, and width, consistent with their stronger couplings, but the impact on the cross section is otherwise similar for both classes. Mirroring the cross section, the $A_{FB}^{*}$ distribution clearly distinguishes between the models and SM, with the difference in width even more readily apparent. The best distinguishing power over all the models investigated comes from the $A_{L}$ distribution, which features an oppositely signed peak for the GLR and GSM classes.

To evaluate the significance of each asymmetry as a combined discovery observable we bin in both $m_{tt}$ and its defining variable. For $A_{FB}^*$, the asymmetry is calculated directly. Therefore, we divide the domain of $\cos\theta^*$ into just two equal regions. $A_L$ is extracted from the gradient of the fit to $\cos\theta_l$ for each mass slice, and we calculate the significance directly from this histogram. The final results of the likelihood-based test, as applied to each model, and tested against the SM, are presented in table~\ref{tab:significance}. The models with non-trivial asymmetries consistently show an increased significance for the 2D histograms compared with using $m_{tt}$ alone, illustrating their potential application in gathering evidence to herald the discovery of new physics.

\begin{table}
  \footnotesize
  \centering
  \begin{tabular}{|llccc|}
    \hline
    \bigstrut
    Class & U$(1)'$ & \multicolumn{3}{c|}{Significance ($Z$)} \\
    \cline{3-5}
    \bigstrut
    & & $m_{tt}$ & $m_{tt}$ \& $\cos\theta^{*}$ & $m_{tt}$ \& $\cos\theta_l$ \\
    \hline
    \bigstrut[t]
    \multirow{6}{*}{E$_6$} & U$(1)_\chi$    & $ 3.7$ & -       & -      \\
                           & U$(1)_\psi$    & $ 5.0$ & -       & -      \\
                           & U$(1)_\eta$    & $ 6.1$ & -       & -      \\
                           & U$(1)_S$       & $ 3.4$ & -       & -      \\
                           & U$(1)_I$       & $ 3.4$ & -       & -      \\
                           & U$(1)_N$       & $ 3.5$ & -       & -      \\
    \hline
    \bigstrut[t]
    \multirow{4}{*}{GLR}   & U$(1)_{R }$    & $ 7.7$ & $ 8.5$  & $ 8.6$ \\
                           & U$(1)_{B-L}$   & $ 3.6$ & -       & -      \\
                           & U$(1)_{LR}$    & $ 5.1$ & $ 5.6$  & $ 5.8$ \\
                           & U$(1)_{Y }$    & $ 6.3$ & $ 6.8$  & $ 7.0$ \\
    \hline
    \bigstrut[t]               
    \multirow{3}{*}{GSM}   & U$(1)_{T^3_L}$ & $12.1$ & $13.0$  & $14.0$ \\
                           & U$(1)_{SM}$    & $ 7.1$ & $ 7.3$  & $ 7.6$ \\
                           & U$(1)_{Q}$     & $24.8$ & -       & -      \\
    \hline
  \end{tabular}
  \caption{Expected significance, expressed as the Gaussian equivalent of the $p$-value.}
  \label{tab:significance}
\end{table}

\section{Conclusions}
\label{sec:conclusions}

We have investigated the scope of the LHC in accessing semileptonic final states produced by $t\bar t$ pairs emerging from the decay of a heavy $Z'$ state. We tested a variety of BSM scenarios embedding one such a state, and show that asymmetry observables can be used to not only aid the diagnostic capabilities provided by the cross section, in identifying the nature of a possible $Z'$ signal, but also to increase the combined significance for first discovery. While the analysis was performed at the parton level, we have implemented a reconstruction procedure of the (anti)tops that closely mimics experimental conditions. We have, therefore, set the stage for a fully-fledged analysis eventually also to include parton-shower, hadronisation, and detector reconstruction, which will constitute the subject of a forthcoming publication. In short, we believe that our results represent a significant phenomenological advancement in proving that charge and spin asymmetry observables can have a strong impact in accessing and profiling $Z'\to t\bar t$ signals during Run 2 of the LHC. This is all the more important in view of the fact that several BSM scenarios, chiefly those assigning a composite nature to the recently discovered Higgs boson, embed one or more $Z'$ state which are strongly coupled to top (anti)quarks~\cite{barducci2012}.

\section*{Acknowledgements}

We acknowledge the support of ERC-CoG Horizon 2020, NPTEV-TQP2020 grant no. 648723, European Union. DM is supported by the NExT Institute and an ATLAS PhD Grant, awarded in February 2015. SM is supported in part by the NExT Institute and the STFC Consolidated Grant ST/L000296/1. FS is supported by the STFC Consolidated Grant ST/K001264/1. We would like to thank Ken Mimasu for all his prior work on $Z'$ phenomenology in $t\bar{t}$, as well as his input when creating the generation tools used for this analysis. Thanks also go to Juri Fiaschi for helping us to validate our tools in the case of Drell-Yan $Z'$ production. Additionally, we are very grateful to Glen Cowan for discussions on the statistical procedure, and Lorenzo Moneta for aiding with the implementation.

\bibliography{references}{}
\bibliographystyle{h-physrev}

\end{document}